\documentclass[JOURNALCODE,twocolumn]{USG}
\usepackage{anyfontsize}

\usepackage{colortbl}
\usepackage{xcolor}

\usepackage{steinmetz}
\usepackage[numbers,sort&compress]{natbib}

\graphicspath{{./images/}}
\begin{document}

\title{Zeitgeist-Aware Multimodal (ZAM) Datasets of Pro-Eating Disorder Short-Form Videos: An Idea Worth Researching}

\author[1]{Eden Shaveet}[https://orcid.org/0000-0002-6889-6064]

\author[2,6]{Zefan Sramek}[https://orcid.org/0000-0003-4541-0399]

\author[4,5]{Yumi Hamamoto}[https://orcid.org/0000-0002-7166-6059]

\author[6]{Jing Du}[https://orcid.org/0000-0003-4113-0875]

\author[7]{Scott Griffiths}[https://orcid.org/0000-0002-6366-3560]

\author[1]{Thalia Zhang}

\author[1]{Thalia Viranda}[https://orcid.org/0009-0008-6933-3926]

\author[8]{William Hornby}

\author[6]{Flora Salim}[https://orcid.org/0000-0002-1237-1664]

\author[2,3]{Koji Yatani}[https://orcid.org/0000-0003-4192-0420]

\author[1]{Tanzeem Choudhury}[https://orcid.org/0000-0002-5952-4955]

\authormark{SHAVEET \textsc{et al.}}
\titlemark{ZAM Datasets}

\address[1]{\orgdiv{Department of Information Science, }\orgname{Cornell University, }%
\orgaddress{\state{New York, }\country{USA}}}

\address[2]{\orgdiv{Department of Electrical Engineering and
Information Systems, }\orgname{The University of Tokyo, }%
\orgaddress{\state{Tokyo, }\country{Japan}}}

\address[3]{\orgdiv{Department of Information and Communication Engineering, }\orgname{The University of Tokyo, }%
\orgaddress{\state{Tokyo, }\country{Japan}}}

\address[4]{\orgdiv{Frontier Research Institute for Interdisciplinary Sciences (FRIS), }\orgname{Tohoku University, }%
\orgaddress{\state{Sendai, }\country{Japan}}}

\address[5]{\orgdiv{Research Institute of Electrical Communication (RIEC), }\orgname{Tohoku University, }%
\orgaddress{\state{Sendai, }\country{Japan}}}

\address[6]{\orgdiv{Department of Computer Science and Engineering, }\orgname{The University of New South Wales (UNSW), }%
\orgaddress{\state{Sydney, }\country{Australia}}}

\address[7]{\orgdiv{Melbourne School of Psychological Sciences, }\orgname{University of Melbourne, }%
\orgaddress{\state{Melbourne, }\country{Australia}}}

\address[8]{\orgname{williamhornby.com, }%
\orgaddress{\country{USA}}}

\corres{Eden Shaveet (\email{ems486@cornell.edu})}





\abstract[ABSTRACT]{\textbf{Objective:} Reliable identification of pro-eating disorder (pro-ED) content online suffers from two pervasive problems: 1) existing methods predominantly rely on text-based signals, failing to capture the inherently multimodal nature of multimedia content; and 2) these methods struggle to keep pace with the rapid evolution of references, memes, terminology, and contextual cues that underlie this content. Together, these limitations point to a gap: the absence of an expert-annotated reference standard capable of supporting real-time research and robust multimodal detection model training for pro-ED content on short-form video platforms. \textbf{Method:} To address this, we propose ``zeitgeist-aware'' multimodal (ZAM) datasets: continuously curated collections of annotated multimodal pro-ED content with inclusion criteria that evolve alongside the memetic zeitgeist: the variable essence of what is considered pro-ED as new media and references come into the cultural zeitgeist and are absorbed and interpreted in online spaces. \textbf{Results:} We present a rationale for such datasets, define their core characteristics, outline approaches for their curation, and describe our progress toward that end. \textbf{Discussion:} This dataset and pipeline architecture may benefit researchers across several fields who are interested in how pro-ED sentiment is encoded and transmitted through short-form video content across time, including for the purpose of responsive moderation efforts.}







\maketitle


\section{Introduction}\label{sec:intro}

The proliferation of pro-eating disorder (pro-ED) content online is a well-documented problem that has continuously outpaced efforts to address it. Broadly, pro-ED content refers to material that promotes, glorifies, or normalizes disordered eating behaviors~\cite{borzekowski_e-ana_2010, custers_urgent_2015, dias_ana_2003, lewis_searching_2012, rouleau_potential_2011}, typically framing restrictive or compensatory practices as lifestyle choices or markers of self-discipline rather than as clinically significant behaviors warranting intervention~\cite{csipke_pro-eating_2007}. The mediums through which such content circulates online, however, are continuously evolving, shaped by emerging platforms with varied communication channels~\cite{custers_urgent_2015}, multimedia formats, deliberate efforts to evade content moderation~\cite{bickham_hidden_2024, chancellor_thyghgapp_2016}, and shifts in cultural discourse that alter the meaning and reception of media. Given this, a key factor hindering the containment, moderation, and measurement of pro-ED content online is the difficulty of identifying or approximating ``ground truth'' for such content in real time. This challenge is compounded by the fact that harm is not uniform: material one user finds benign may, in context, promote disordered eating behaviors in another. Exercise-related content, for instance, may worsen body-image disturbances in one viewer while remaining innocuous to another~\cite{capri_differences_2026, yee_differential_2020}. Importantly, content does not need to be explicitly pro-ED in intent to exacerbate ED symptoms. Of primary concern here, however, is the narrower category of material that explicitly glorifies disordered eating behaviors, is increasingly difficult to detect across multimedia formats, and whose pro-ED intent is most legible to those familiar with in-group signaling conventions, yet coded enough to evade traditional moderation systems.
The rise of short-form video platforms, such as TikTok, Instagram Reels, and YouTube Shorts, has introduced new complexities in the detection of this sort of pro-ED material. Content moderation pipelines have historically favored text-based identification techniques, such as the detection and banning of specific terms or hashtags, developed during an era of text-oriented social media in which visual elements served as supplements to text rather than as the primary modality of information delivery~\cite{chancellor_multimodal_2017}. These approaches have proven inadequate in the short-form video landscape, where pro-ED sentiment is communicated through the confluence and layering of multiple modalities (visual, auditory, and textual) into memetic formats~\cite{bickham_edtok_2025, chu_bigtokdetect_2026, tao_unraveling_2026}.
To address this problem, we propose zeitgeist-aware multimodal (ZAM) datasets, continuously curated collections of multimodally anatomized pro-ED videos (videos examined by their modal components) whose inclusion criteria evolve alongside the memetic zeitgeist, or the evolving essence of what is considered pro-ED as new media and references come into the cultural zeitgeist and are absorbed and interpreted in online spaces. In this paper, we provide a rationale for such datasets, define their key characteristics, propose approaches for their creation, and describe our progress.

\section{Background}
\label{sec:background}

\subsection{History of Pro-ED Content Detection}

The first documented examples of explicitly pro-ED content online appeared on dedicated webpages hosted by free web hosting services in the 1990s and early 2000s~\cite{dias_ana_2003}. These pages prominently featured images of extremely thin bodies alongside aspirational text promoting thin body ideals, a genre of content most commonly referred to as ``thinspiration'' or ``thinspo''~\cite{borzekowski_e-ana_2010, dias_ana_2003}. Initial efforts to moderate this content took the form of manual webpage removal, prompted by complaints from eating disorder advocacy organizations and media exposés~\cite{holahan_yahoo_2001, winfrey_girls_2001}. With the emergence of the first social media platforms in the early 2000s, pro-ED content migrated to these new spaces, and individual platforms were obliged to develop their own detection and moderation approaches. Both web hosting and emerging social media platforms justified removals by citing violations of terms of service related to self-harm~\cite{george_nurturing_2002, staff_follow-up_1201}, relying on human review, keyword searches, and complaints to identify pro-ED content. However, this approach was difficult to sustain as platform user bases grew rapidly~\cite{gillespie_content_2020}. As users dispersed across an expanding collection of platforms, content detection shifted toward automated, text-based approaches that drew on keyword filtering, hashtag analysis, and algorithmic classification~\cite{chancellor_multimodal_2017}. This reliance continued even as social media platforms trended toward multimodal communication, for which text is not always the primary conveyor of meaning~\cite{chu_bigtokdetect_2026}. The same pattern held for research on pro-ED social media content. Research applying computational methods to pro-ED content detection has been predominantly text-based~\cite{merhbene_investigating_2024}, and studies incorporating visual content often still depend on text-based signals (such as hashtags) for initial data collection~\cite{chancellor_multimodal_2017, chu_bigtokdetect_2026}, leaving textless or primarily audio-visual content underrepresented.

\subsection{Short-Form Video \& Multimodal Pro-ED Content}

Challenges in content identification have been compounded by the rise of short-form video (SFV) platforms. SFV platforms employ recommendation-driven feeds, such as TikTok’s ``For You Page,'' that dynamically adapt to user interactions, generating personalized content streams that incentivize rapid, passive consumption. Recent work has estimated that people with eating disorders are algorithmically delivered over 4,000\% more pro-ED videos on TikTok than those without eating disorders~\cite{griffiths_does_2024}. This may be largely driven by platform recommendation systems that exploit user preferences to sustain engagement, even when the resulting content stream may not benefit users, or may even harm them through the nature of the content itself or through compulsive prolonged use~\cite{jain_exploring_2025, metzler_social_2024}. The SFV format has been widely adopted across major social media platforms, emerging as a dominant mode of digital content consumption. As multimedia content proliferates across these platforms, effective moderation has become increasingly dependent on the ability to reliably detect and identify pro-ED content through automated means, and has grown substantially more difficult as social media has shifted from predominantly text-based platforms to those centered on video and multimedia.

On video platforms, pro-ED sentiment is typically constructed through multimodal confluence: the co-occurrence and interaction of visual, auditory, textual, and memetic elements, whose meanings evolve over time. To illustrate this, consider a short-form video featuring imagery of an empty bowl. In isolation, such imagery may be entirely innocuous. However, when overlaid with a low-tempo song whose lyrics reference wanting a ``perfect body,'' its valence shifts. Consider the addition of on-screen text that simply reads ``breakfast'' and a single hashtag in the description: ``\#3D.'' Within the eating disorder community, ``3D'' is a coded reference to ``ED.'' Unlike more explicit tags such as ``thinspo'' or its variants (e.g., ``th!nsp0''), it cannot be uniformly blocked because it carries harmless meanings in other contexts, such as in three-dimensional modeling. As demonstrated in Figure 1, each of these elements is benign when examined individually; however, when interpreted in combination, pro-ED sentiment becomes clearly discernible. While multimodal sentiment exists across video platforms, the digital affordances of SFV platforms, including low-friction production and consumption, compressed content lifecycles, and algorithmic feed tailoring, amplify its reach.

\begin{figure*}[htb]
  \centering
  \includegraphics[width=\textwidth]{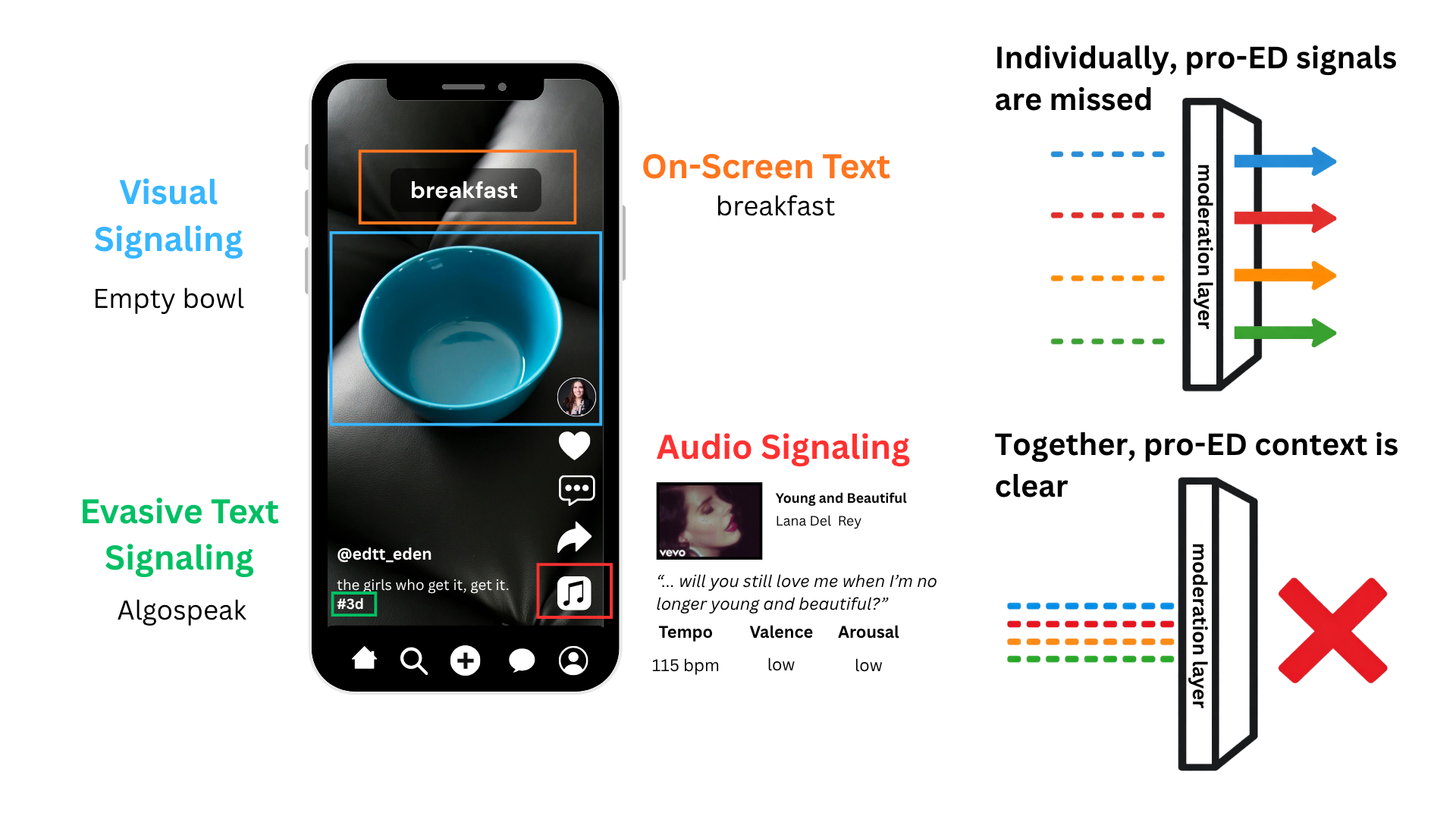}
  \caption{Example evasive pro-eating disorder short-form video signal processing when signals are evaluated independently vs together.\label{fig:fig1}}
\end{figure*}

\subsection{Evasive Patterns in Pro-ED Content}
Exploitation of the limited nature of text-based detection methods is especially salient in content deliberately designed to evade moderation through devices such as ``algospeak,'' the substitution or obfuscation of flagged terms using alternative characters (e.g., ``th!nsp0'' as algospeak for ``thinspo'')~\cite{chancellor_thyghgapp_2016, steen_you_2023}, or coded in-group visual references, such as depictions of deer and fawns used to romanticize frailty and malnourishment. In some content, such imagery is accompanied by the word ``accountability,'' which signals to in-group members that they should report their height and weights in the comments (Figure 2a). In other content, deer imagery is overlaid by on-screen text that hints at disordered eating behaviors and the subtext of being ``a deer in the headlights'' (Figure 2b-c). These signals convey meaning to community members while remaining opaque to outsiders~\cite{gerrard_beyond_2018}. The evasiveness of algospeak is a self-compounding problem. Text-based moderation typically operates by uniformly banning specific terms, and algospeak terms are formed to circumvent these bans~\cite{steen_you_2023}. When these variants are themselves banned, new alternatives are continuously generated, or existing innocuous terms are co-opted (such as ``\#3D'' or ``\#edsheeran''~\cite{greene_visions_2023}) precisely because they cannot be blocked without prohibiting innocuous use. This cycle persists as users communicate through non-textual signals, such as visual and auditory, that operate alongside or in place of text, further undermining text-dependent moderation. Multimodality has compounded this challenge. Text embedded within images or video that corresponds in some way to the audio playing in the background (such as modified lyrics to express pro-ED sentiment) is often easily detected by human viewers but remains difficult for automated systems to identify, interpret, or evaluate against a rapidly shifting cultural context.

\subsection{Memetic Zeitgeist of Pro-Eating Disorder Content}
A second prevailing issue in the detection of pro-ED content is the fact that what is considered explicitly ``pro-ED'' changes rapidly over time as in-group language, coded use of sound, and imagery evolves. Things that were once considered innocuous may become associated with pro-ED sentiment after cultural events and phenomena, and pro-ED sentiment evolves as new media and references come into the cultural zeitgeist. These references are absorbed and appropriated, creating a parallel ``memetic zeitgeist'' that is only known to in-group members, and that we posit must be known or approximated in order to understand to what extent content is explicitly pro-ED. For instance, the recent popularity of ``K-pop'' (Korean popular music) has given rise to a distinct form of pro-ED content grounded in body-image ideals of K-pop stars (called ``idols''), who are closely monitored and expected to maintain weight ranges set by their management agencies via intensive weight loss and calorie restriction regimens~\cite{venters_trammelled_2023}. In these videos, K-pop idols are shown either statically (as an image) or dynamically (dancing), often with overlaid text referencing a user’s weight-related metrics or masked references to desired compensatory behaviors (Figures 2d-g). Prior to the early 2020s, K-pop imagery was less commonly associated with pro-ED content and would have had a less salient presence in such a dataset. Such videos have since proliferated however, and their representation in a dataset drawn from a later period would be substantially greater. The relevance of pro-ED signals ebbs and flows, with temporality playing a key role relative to memetic zeitgeist.

\begin{figure*}[htb]
  \centering
  \includegraphics[width=\textwidth]{images/ijed_fig2.png}
  \caption{2a-c: Example evasive deer-oriented ED content on BlueSky and TikTok. Figure 2a is a BlueSky post featuring an image of a girl sitting with a fawn, while the text reads ``accountability thread.'' Figure 2b is a still from a TikTok video featuring a deer with overlaid, barely visible text: ``The guilt after eating.'' Figure 2c is a still from a TikTok video featuring a deer with overlaid text, in quotations: ``‘you need to eat something’.'' 2e-f: Example K-Pop-oriented ``What I Eat In a Day'' (wieiad) TikTok posts featuring photos of K-pop idols (Left: Jiwoo; Right: Jonghyun). Both Figures 2e and 2f feature K-pop star imagery in their first still and consumption-related metrics in their last. 2d \& 2g: Example TikTok video stills featuring dancing K-Pop idols with overlaid pro-ED messaging on-screen. Figure 2d features imagery of K-pop idol, Yuna, dancing, while on-screen text reads ``Keep on [star emoji]ving,'' as an evasive way to say ``starving.'' Figure 2g features imagery of K-Pop idol, Moka, dancing while on-screen text reads ``if you start now you’ll have your dream body by summer.''\label{fig:fig2}}
\end{figure*}

\section{An Idea Worth Researching: Zeitgeist-Aware Multimodal (ZAM) Datasets}
As discussed, reliable identification of pro-eating disorder content online suffers from two pervasive problems: 1) existing identification methods overwhelmingly rely on text-based indicators, which are insufficient in the age of pervasive multimedia in which signals are multimodal; 2) the references, memes, terms, and contextual flags which underlie this content evolve rapidly over time. Given this, a prevailing problem in research and moderation around pro-ED content alike is the lack of existence of an expert-annotated reference standard dataset that may be used to guide and inform real-time research related to pro-ED content on short-form video platforms, including training and fine-tuning robust multimodal content detection models. To address this gap, we propose the development of zeitgeist-aware multimodal (ZAM) datasets: continuously curated collections of multimodally anatomized pro-ED videos whose inclusion criteria evolve with the memetic zeitgeist.


The curation of ZAM datasets of pro-ED content will require robust systems that account for several factors, including multimodality, memetic zeitgeist, and clinical relevance. We envision this system comprising two interrelated components operating within a single pipeline: 1) human experts in the loop, and 2) low-friction interactive systems. Humans-in-the-loop would draw on two complementary forms of expertise. The first is that of zeitgeist-aware individuals with current, firsthand knowledge of pro-ED content. These may be individuals recovering from body-image disturbance or disordered eating who encounter unwanted pro-ED material during routine platform use. The second is that of clinical experts specializing in disordered eating and body image disturbance, equipped to assess whether content promotes, glorifies, or normalizes clinically relevant ED behaviors that reflect eating disorder psychopathology, such as those derived from the Eating Disorder Examination (EDE) and diagnostic items from the Diagnostic and Statistical Manual of Mental Disorders (DSM)~\cite{cooper_eating_2011, noauthor_diagnostic_2013}, informed by contextual insights from zeitgeist-aware collaborators. These two forms of expertise are interdependent in the context of data curation and annotation. In the case of evasive pro-ED content, for instance, only zeitgeist-aware individuals would know how to interpret in-group signaling, while clinicians would have the clinical grounding to assess whether the interpreted meaning promotes clinically significant behaviors. In the curation of ZAM datasets, both forms of expertise are complementary and necessary.
In turn, the low-friction interactive systems would provide a way for zeitgeist aware individuals to contribute data in the digital spaces where they already are. We argue that the problem of identifying pro-ED content at scale can be effectively addressed by tapping into the network of zeitgeist-aware users who are already participating in these online spaces. To facilitate this, a purpose-designed low-friction interactive data contribution system would make contributing to the dataset seamless, requiring little to no difference in how individuals are already interacting with their short-form video platforms, making use of their existing interaction flows. Additionally, separate annotation interfaces would be accessible to annotators to validate automatically extracted information (e.g., confirming that transcribed audio and detected objects are correct) and annotated for pro-ED sentiment and features. Together, this information would be coalesced and processed to anatomize multimodal features alongside clinical and memetic annotations.
Toward this goal, we have made meaningful progress in building a ZAM curation system. Our work began with the development of EDTT, a small but rigorously annotated dataset of pro-eating disorder content drawn from TikTok~\cite{shaveet_edtt_2026}. Clinically grounded and multimodal in nature, it features inductively derived human annotations spanning visual, audio, and pro-ED sentiment dimensions, produced through cross-disciplinary collaboration, but only for a one-year period. Building on this foundation, we are prototyping a ZAM-curation pipeline intended to operate on a recurring basis, drawing inspiration from EDTT while expanding its field coverage using methods including transcription, object detection, optical character recognition, and audio processing, while maintaining responsiveness to the evolving pro-ED memetic zeitgeist. This dataset and pipeline architecture may benefit researchers across computer science, sociology, public health, and related fields who are interested in how pro-ED sentiment is encoded and transmitted through short-form video content across time, including for the purpose of responsive moderation efforts.

\section{Conclusion}\label{sec:conclusions}

As pro-ED content continues to evolve and proliferate across multimedia platforms, our efforts to identify it, for both moderation and research purposes, must evolve alongside it. Left unaddressed, existing content detection strategies will continue to fail short-form video platform users who have, or are vulnerable to, eating disorders. Additionally, peer-reviewed research on pro-ED content, such as thematic analysis, carries conception-to-publication timelines that can exceed one-to-two years, leaving the field perpetually behind the content actually reaching users. A reliable ZAM dataset curation system would address both problems, enabling more effective content detection and removal, as well as empirical research that is timely enough to be practical. Building on existing multimodal content annotation work, we aim to contribute to this space through the development of our ZAM-curation pipeline, designed in partnership with the communities this work seeks to serve: individuals who wish to limit their exposure to pro-ED content on short-form video platforms, and clinicians seeking the same for their patients. We believe ZAM datasets represent a promising and underexplored idea worth researching.

\bmsubsection*{Author Contributions}

\textbf{Eden Shaveet:} conceptualization, supervision, investigation, visualization, methodology, project administration, writing - original draft
\textbf{Zefan Sramek:} conceptualization, methodology, investigation, writing - review \& editing
\textbf{Yumi Hamamoto:} conceptualization, methodology, investigation, writing - review \& editing
\textbf{Jing Du:} conceptualization, methodology, investigation, writing - review \& editing
\textbf{Scott Griffiths:} writing - review \& editing
\textbf{Thalia Zhang:} writing - review \& editing
\textbf{Thalia Viranda:} writing - review \& editing
\textbf{William Hornby:} writing - review \& editing
\textbf{Flora Salim:} writing - review \& editing
\textbf{Koji Yatani:} writing - review \& editing, funding acquisition
\textbf{Tanzeem Choudhury:} writing - review \& editing, funding acquisition

\bmsubsection*{Acknowledgments}

We are grateful to the ASPIRE-Mental Well-being Intelligence (MWI) initiative and community for bringing several of the authors together.

\bmsubsection*{Financial Disclosure}

This research is supported by the Japan Science and Technology Agency's JST ASPIRE for Top Scientists (Grant Number: JPMJAP2405), and by the ARC Centre of Excellence for Automated Decision-Making and Society (No. CE200100005), funded by the Australian Government through the Australian Research Council.
Eden Shaveet is supported by a National Science Foundation CISE Graduate Fellowship (CSGrad4US) under Grant No. 2313998. Any opinions, findings, and conclusions or recommendations expressed in this material are those of the authors and do not necessarily reflect the views of the National Science Foundation.

\bmsubsection*{Conflicts of Interest}

Scott Griffiths has provided unpaid expertise to government bodies in Australia, Singapore, and the United Kingdom; unpaid expertise to Meta; paid expertise to TikTok; and a combination of paid and unpaid expertise to education- and mental health-related organizations in Australia. Tanzeem Choudhury holds equity stake in Dapple Health and SomaBeat Health.

\bibliographystyle{unsrt}
\bibliography{IJED}



\end{document}